\documentclass[12pt,english]{article}
\usepackage[T1]{fontenc}
\usepackage[latin1]{inputenc}
\usepackage{amsmath}
\usepackage{amssymb}
\usepackage{esint}

\makeatletter
\usepackage{babel}\usepackage{amsfonts}
\makeatletter
\providecommand{\LyX}{L\kern-.1667em\lower.25em\hbox{Y}\kern-.125emX\@}
\providecommand{\LyX}{L\kern-.1667em\lower.25em\hbox{Y}\kern-.125emX\@}
\newtheorem{theorem}{Theorem}

\newtheorem{lemma}[theorem]{Lemma}

\makeatother

\makeatother

\usepackage{babel}

\begin{document}

\title{On Large Densities in Fermi Systems}

\author{V. A. Malyshev \and A. A. Zamyatin}

\date{}
\maketitle
\begin{abstract}
The goal of this paper is to give some rigorous results, concerning
high density behavior of Fermi systems.

\end{abstract}

\section{Introduction}

The goal of this paper is to prove some rigorous results, concerning
high density behavior of Fermi systems. This behavior had been presented
in Landau-Lifshitz course, see p. 345 of \cite{LanLif}, as follows:
for large densities the Fermi system becomes ideal, they mean by this
that the kinetic energy of most particles becomes much greater than
their interaction (with other particles or external field) energy.
However, some work is necessary to make this clear physical picture
mathematically well-defined.

It is well known that in the equilibrium quantum statistical mechanics,
as $\beta\rightarrow\infty$ and under some mild conditions, $\beta$-KMS
state tends to the ground state, see \cite{BraRob}. On the contrary
for fixed $\beta$, if the chemical potential $\mu$ (for grand canonical
ensemble) or the density $\rho$ of particles (for canonical ensemble)
tend to infinity, there is no natural limiting state. In this case
it is reasonable to study the large $\mu$ or large $\rho$ asymptotics
of thermodynamic functions, in particular of $\ln Z$, where $Z$
is the partition function.

We want to note that heuristic arguments for the classical system
show that it is only possible when the interaction tends to zero as
the volume and density tend to infinity. In fact, take the volume
$\Lambda$ with $\rho\Lambda$ particles. Let the radius of potential
be $r$, assume it bounded by $v$ and local. Introduce external field
so that this mean interaction energy were $0$, in other words, that
the system were neutral. By central limit theorem for most particles
the interaction energy is of order $v\sqrt{\rho}r^{\frac{d}{2}}$.
At the same time the kinetic energy is of order $c(\beta)$. Thus
the kinetic energy exceeds the interaction energy if only $vr^{\frac{d}{2}}=o(\rho^{-\frac{1}{2}})$,
or $o(\rho^{-1})$ without neutrality condition.

Our rigorous results show that in an one-dimensional model the smallness
of potential is not really necessary. This is not quite evident because
standard spectrum perturbation theory is not applicable, \ since
difference $\varepsilon_{k}-\varepsilon_{k-1}$ between consecutive
eigenvalues tends to zero, if $k\ll\Lambda^{2}.$

In \cite{LanLif} neutral systems of $M$ nuclei and $N=qM$ electrons
are considered, where $q$ is the charge of the nuclei. Assuming that
the nuclei are situated in the vertices of a periodic lattice and
the electrons do not interact with each other, we model this as the
system of independent spinless Fermi particles in the external potential
$V$.

\section{Results}

Now we give exact definitions. In a finite interval $\Lambda\subset R$
we consider (one-particle) Hamiltonian\[
h=-\nabla^{2}+V\]
 in $L^{2}(\Lambda)$ with Dirichlet boundary conditions, but other
boundary conditions could be considered as well. $V$ is the multiplication
operator on the function $V(x)$, it is always assumed to be bounded.
Let\[
\varepsilon_{1}<\varepsilon_{2}<...\,\varepsilon_{k}=\varepsilon_{k}^{(V)}(\Lambda)<...\]
 be the eigenvalues of this one particle problem. It is known that
there are no multiple eigenvalues, see \cite{Danf}, theorem 13.7.50.
It is known also that for $V=0$\[
\varepsilon_{k}^{(0)}(\Lambda)=\pi^{2}\frac{k^{2}}{\Lambda^{2}}\]

Now consider the Fermi gas of independent particles in the external
field $V$ in $\Lambda$. The canonical partition function is\[
Z_{N}^{(V)}(\Lambda)=\sum_{0<k_{1}<k_{2}<...<k_{N}}\exp\left(-\beta\sum\limits _{i=1}^{N}\varepsilon_{k_{i}}^{(V)}(\Lambda)\right),\]
 where $N$ is the number of particles and $\Lambda$ is the \textquotedbl{}volume\textquotedbl{}.
We always assume that the potential $V$ is such that the limit\[
F=F(\beta,\rho)=\lim\frac{1}{\Lambda}\ln Z_{N}^{(V)}(\Lambda)\]
 exists if $\Lambda\rightarrow\infty$ so that $\frac{N}{\Lambda}=\rho$.

\begin{theorem} Let $V$ be bounded potential. Then as $N$ and $\Lambda$
tend to infinity so that $\frac{N}{\Lambda}\rightarrow\infty$ we
have \[
\frac{\ln Z_{N}^{(V)}(\Lambda)}{\ln Z_{N}^{(0)}(\Lambda)}\rightarrow1\]

\end{theorem}

The grand canonical partition function for Fermi gas in external field
with potential $V$ is\[
\Xi_{\mu}^{(V)}(\Lambda)=\prod\limits _{k=1}^{\infty}\left(1+ze^{-\beta\varepsilon_{k}^{(V)}(\Lambda)}\right),\]
 where $z=\exp(\mu\beta)$ and $\mu$ is the chemical potential.

\begin{theorem} Let $\Lambda$ be fixed and $V$ be bounded. Then
as $\mu\rightarrow\infty$\[
\frac{\ln\Xi_{\mu}^{(V)}\left(\Lambda\right)}{\ln\Xi_{\mu}^{(0)}\left(\Lambda\right)}\rightarrow1\]

\end{theorem}

Put\[
\Omega_{\mu}^{(V)}(\Lambda)=\frac{1}{\Lambda}\ln\Xi_{\mu}^{(V)}\left(\Lambda\right)\]
 We always assume that the potential $V$ is such that the limit\[
\Omega_{\mu}^{(V)}=\lim_{\Lambda\rightarrow\infty}\frac{1}{\Lambda}\ln\Xi_{\mu}^{(V)}\left(\Lambda\right)\]
 exists.

\begin{theorem} As $\mu\rightarrow\infty$\[
\frac{\Omega_{\mu}^{(V)}}{\Omega_{\mu}^{(0)}}\rightarrow1\]

\end{theorem}

\section{Proofs}

We will need the following result.

\begin{lemma} If $V$ is bounded then for any $\Lambda$\[
\left\vert \varepsilon_{k}(\Lambda,V)-\varepsilon_{k}(\Lambda,0)\right\vert <C\]

\end{lemma}

\noindent \textbf{Proof.} This fact can be found in the proof of theorem
13.82$\frac{1}{2}$ in \cite{ReeSim4}. We give the sketch of the
proof. Consider the analytic family of operators (for fixed $\Lambda$)
\[
H(a)=-\nabla^{2}+aV,0\leq a\leq1\]
 From simplicity of the eigenvalues for any $a$ one can deduce that
they analytically depend on $a$ on this interval. Then using the
formula\[
\frac{d\varepsilon_{k}(a)}{da}=(\phi_{k}(a),aV\phi_{k}(a))\]
 where $\phi_{k}(a)$ are the corresponding eigenvectors with norm
$1$, we get\[
\left\vert \varepsilon_{k}(1)-\varepsilon_{k}(0)\right\vert \leq\sup_{\lbrack0,\Lambda]}\left\vert V\right\vert =C\]

\subsection{Canonical ensemble}

\begin{lemma} Let $N>\Lambda.$ Then \begin{equation}
Z_{N}^{(0)}(\Lambda)\leq\frac{1}{N!}\left(\frac{\Lambda}{\beta}\right)^{N}\left(1+\frac{e^{\beta}\beta N}{\Lambda}\right)^{\Lambda}\label{est1}\end{equation}

\end{lemma}

\noindent \textbf{Proof.} Set for $p=1,...,\left[\Lambda\right]$\begin{multline*}
Z_{N,p}(\Lambda)=\sum_{\substack{0<k_{1}<...<k_{p}\leq\left[\Lambda\right]\\
<k_{p+1}<...<k_{N}}
}\exp\left(-\beta\sum\limits _{i=1}^{N}\varepsilon_{k_{i}}^{(0)}(\Lambda)\right)=\\
\sum_{0<k_{1}<...<k_{p}\leq\left[\Lambda\right]}\exp\left(-\beta\sum\limits _{i=1}^{p}\varepsilon_{k_{i}}^{(0)}(\Lambda)\right)\sum_{\left[\Lambda\right]<k_{p+1}<...<k_{N}}\exp\left(-\beta\sum\limits _{i=p+1}^{N}\varepsilon_{k_{i}}^{(0)}(\Lambda)\right)\end{multline*}
 and for $p=0$\[
Z_{N,0}(\Lambda)=\sum_{\left[\Lambda\right]<k_{1}<...<k_{N}}\exp\left(-\beta\sum\limits _{i=1}^{N}\varepsilon_{k_{i}}^{(0)}(\Lambda)\right)\]
 Then\[
Z_{N}^{(0)}(\Lambda)=\sum_{p=0}^{\left[\Lambda\right]}Z_{N,p}(\Lambda)\]
 If $k_{i}>\left[\Lambda\right],$ then $\varepsilon_{k_{i}}^{(0)}(\Lambda)>\frac{k_{i}}{\Lambda}$
and\begin{multline*}
\sum_{\left[\Lambda\right]<k_{p+1}<...<k_{N}}\exp\left(-\beta\sum\limits _{i=p+1}^{N}\varepsilon_{k_{i}}^{(0)}(\Lambda)\right)\leq\sum_{\left[\Lambda\right]<k_{p+1}<...<k_{N}}\exp\left(-\beta\sum\limits _{i=p+1}^{N}\frac{k_{i}}{\Lambda}\right)=\\
\sum_{\left[\Lambda\right]<k_{p+1}}\exp\left(-\frac{\beta k_{p+1}}{\Lambda}\right)...\sum_{k_{N-1}<k_{N}}\exp\left(-\frac{\beta k_{N}}{\Lambda}\right)=\\
\exp\left(-\left(N-p\right)\beta\right)\prod\limits _{k=1}^{N-p}\frac{\exp\left(-\frac{\beta k}{\Lambda}\right)}{1-\exp\left(-\frac{\beta k}{\Lambda}\right)}=\exp\left(-\left(N-p\right)\beta\right)\prod\limits _{k=1}^{N-p}\frac{1}{\exp\left(\frac{\beta k}{\Lambda}\right)-1}\end{multline*}
 Using the evident estimate $\exp\left(\frac{\beta k}{\Lambda}\right)-1>\frac{\beta k}{\Lambda}$
we get\[
\sum_{\left[\Lambda\right]<k_{p+1}<...<k_{N}}\exp\left(-\beta\sum\limits _{i=p+1}^{N}\varepsilon_{k_{i}}^{(0)}(\Lambda)\right)\leq\frac{\exp\beta\left(p-N\right)}{\left(N-p\right)!}\left(\frac{\Lambda}{\beta}\right)^{N-p}\]
 Further, we may estimate the following sum\[
\sum_{0<k_{1}<...<k_{p}\leq\left[\Lambda\right]}\exp\left(-\beta\sum\limits _{i=1}^{p}\varepsilon_{k_{i}}^{(0)}(\Lambda)\right)\leq\binom{\left[\Lambda\right]}{p}\]
 Hence,\begin{multline*}
Z_{N}^{(0)}(\Lambda)=\sum_{p=0}^{\left[\Lambda\right]}Z_{N,p}(\Lambda)\leq\exp(-\beta N)\sum_{p=0}^{\left[\Lambda\right]}\binom{\left[\Lambda\right]}{p}\frac{\exp\beta p}{\left(N-p\right)!}\left(\frac{\Lambda}{\beta}\right)^{N-p}=\\
\frac{\exp(-\beta N)}{N!}\sum_{p=0}^{\left[\Lambda\right]}\binom{\left[\Lambda\right]}{p}\exp\beta p\frac{N!}{\left(N-p\right)!}\left(\frac{\Lambda}{\beta}\right)^{N-p}\end{multline*}
 But $\frac{N!}{\left(N-p\right)!}\leq N^{p}$ and\begin{align*}
Z_{N}^{(0)}(\Lambda) & \leq\frac{\exp(-\beta N)}{N!}\left(\frac{\Lambda}{\beta}\right)^{N}\sum_{p=0}^{\left[\Lambda\right]}\binom{\left[\Lambda\right]}{p}\exp\beta p\left(\frac{\beta N}{\Lambda}\right)^{p}\leq\\
 & \frac{1}{N!}\left(\frac{\Lambda}{\beta}\right)^{N}\left(1+\frac{e^{\beta}\beta N}{\Lambda}\right)^{\Lambda}\end{align*}
 The lemma is proved.

Take the logarithm of both sides of inequality (\ref{est1}) \ and
divide by $N:$\[
\frac{\ln Z_{N}^{(0)}(\Lambda)}{N}\leq\ln\Lambda-\frac{\ln N!}{N}+\frac{\Lambda}{N}\ln\left(1+\frac{e^{\beta}\beta N}{\Lambda}\right)-\ln\beta\]
 Since for large $N$\[
\ln N!>N\ln N-N\]
 then\begin{align*}
\frac{\ln Z_{N}^{(0)}(\Lambda)}{N} & \leq\ln\Lambda-\ln N+\frac{\Lambda}{N}\ln\left(1+\frac{e^{\beta}\beta N}{\Lambda}\right)-\ln\beta+1=\\
 & \ln\frac{\Lambda}{N}+\frac{\ln\left(1+\frac{e^{\beta}\beta N}{\Lambda}\right)}{\frac{N}{\Lambda}}-\ln\beta+1\end{align*}
 So, we see that the right side of the above inequality tends to $-\infty$
under the conditions of the theorem. It follows that as $N,\Lambda,\frac{N}{\Lambda}\rightarrow\infty$
\begin{equation}
\frac{\ln Z_{N}^{(0)}(\Lambda)}{N}\rightarrow-\infty\label{lim1}\end{equation}
 Return to the proof of the theorem. We need to show that as $N,\Lambda,\frac{N}{\Lambda}\rightarrow\infty$
\begin{equation}
\left\vert \frac{\ln Z_{N}^{(V)}(\Lambda)-\ln Z_{N}^{(0)}(\Lambda)}{\ln Z_{N}^{(0)}(\Lambda)}\right\vert \rightarrow0\label{lim2}\end{equation}
 It follows from lemma 1 that\[
\left\vert \frac{\ln Z_{N}^{(V)}(\Lambda)-\ln Z_{N}^{(0)}(\Lambda)}{\ln Z_{N}^{(0)}(\Lambda)}\right\vert \leq\left\vert \frac{CN}{\ln Z_{N}^{(0)}(\Lambda)}\right\vert \]
 Now (\ref{lim2}) follows from (\ref{lim1}).

\subsection{Grand canonical ensemble}

\paragraph{Proof of theorem 2}

Define distribution function of eigenvalues\[
F_{\Lambda}^{(V)}(t)=\#\{k:\varepsilon_{k}^{(V)}(\Lambda)\leq t\}\]
 We have\begin{equation}
\ln\Xi_{\mu}^{(V)}(\Lambda)=\int_{0}^{\infty}\ln\left(1+e^{\beta(\mu-t)}\right)dF_{\Lambda}^{(V)}(t)\end{equation}
 Let us prove the following lemmas.

\begin{lemma} As $t\rightarrow\infty$ \begin{equation}
F_{\Lambda}^{(V)}(t)=\frac{\Lambda\sqrt{t}}{\pi}+O(1)\label{res}\end{equation}

\end{lemma}

\noindent \textbf{Proof.} We give elementary proof, see general result
in \cite{Shubin}, \cite{Mar}, \cite{SafVas}. Let\[
\varepsilon_{k}^{(V)}(\Lambda)\leq t<\varepsilon_{k+1}^{(V)}(\Lambda)\]
 For such $t:F_{\Lambda}^{(V)}(t)$ $=k.$ Using the estimate (lemma
4)\[
|\varepsilon_{k}^{(V)}(\Lambda)-\varepsilon_{k}^{(0)}(\Lambda)|\leq C\]
 find\[
-C+\frac{k^{2}\pi^{2}}{\Lambda^{2}}\leq t\leq C+\frac{(k+1)^{2}\pi^{2}}{\Lambda^{2}}\]
 After multiplying by $\Lambda^{2}/$ $\pi^{2}$ and extracting the
square root \ we come to the inequality: \[
-1+\frac{\Lambda\sqrt{t}}{\pi}\sqrt{1-\frac{C}{t}}\leq k\leq\frac{\Lambda\sqrt{t}}{\pi}\sqrt{1+\frac{C}{t}}\]
 But $k=F_{\Lambda}^{(V)}(t),$ so\[
-1+\frac{\Lambda\sqrt{t}}{\pi}\sqrt{1-\frac{C}{t}}\leq F_{\Lambda}^{(V)}(t)\leq\frac{\Lambda\sqrt{t}}{\pi}\sqrt{1+\frac{C}{t}}\]

Subtracting $\frac{\Lambda\sqrt{t}}{\pi}$ \ we get \[
-1+\frac{\Lambda\sqrt{t}}{\pi}\left(\sqrt{1-\frac{C}{t}}-1\right)\leq F_{\Lambda}^{(V)}(t)-\frac{\Lambda\sqrt{t}}{\pi}\leq\frac{\Lambda\sqrt{t}}{\pi}\left(\sqrt{1+\frac{C}{t}}-1\right)\]
 For $t$ \ large enough \[
-1-\frac{\Lambda\sqrt{t}}{\pi}\frac{C}{2t}\leq F_{\Lambda}^{(V)}(t)-\frac{\Lambda\sqrt{t}}{\pi}\leq\frac{\Lambda\sqrt{t}}{\pi}\frac{C}{2t}\]
 It gives the result of the lemma.

\begin{lemma} As \bigskip{}
$\mu\rightarrow\infty$ \begin{equation}
\ln\Xi_{\mu}^{(V)}(\Lambda)\sim\frac{2}{3}\frac{\Lambda}{\pi}\mu^{3/2}\label{equiv}\end{equation}

\end{lemma}

\noindent \textbf{Proof.} Indeed,\[
\ln\Xi_{\mu}^{(V)}(\Lambda)=\int_{0}^{\infty}\ln\left(1+e^{\beta(\mu-t)}\right)dF_{\Lambda}^{(V)}(t)\]
 Integrating by parts and using $F_{\Lambda}^{(V)}(t)\sim\frac{\Lambda}{\pi}\sqrt{t}$\ we
have\begin{align*}
\ln\Xi_{\mu}^{(V)}(\Lambda) & =\beta\int_{0}^{\infty}\frac{e^{\beta(\mu-t)}}{1+e^{\beta(\mu-t)}}F_{\Lambda}^{(V)}(t)dt=\\
=\beta\int_{0}^{\mu}F_{\Lambda}^{(V)}(t)dt-\beta\int_{0}^{\mu}\frac{F_{\Lambda}^{(V)}(t)}{1+e^{\beta(\mu-t)}}dt & +\beta\int_{\mu}^{\infty}\frac{F_{\Lambda}^{(V)}(t)}{1+e^{\beta(t-\mu)}}dt\end{align*}
 It follows from (\ref{res}), that\begin{equation}
\int_{0}^{\mu}F_{\Lambda}^{(V)}(t)dt\sim\frac{2}{3}\frac{\Lambda}{\pi}\mu^{3/2}\label{0int}\end{equation}
 Let us show that\begin{equation}
\int_{0}^{\mu}\frac{F_{\Lambda}^{(V)}(t)}{1+e^{\beta(\mu-t)}}dt=O(\mu^{1/2+\alpha}),\label{1int}\end{equation}
 where $\alpha>0$ is arbitrary small. By (\ref{res}) we can write
\[
F_{\Lambda}^{(V)}(t)=c\sqrt{t}+r(t),r(t)=O(1)\]
 Let us prove\[
\int_{0}^{\mu}\frac{\sqrt{t}}{1+e^{\beta(\mu-t)}}dt=O(\mu^{1/2})\]
 After change of variable $s=t-\mu$\[
\int_{0}^{\mu}\frac{\sqrt{t}}{1+e^{\beta(\mu-t)}}dt=\int_{-\mu}^{0}\frac{\sqrt{\mu+s}}{1+e^{-\beta s}}dt=\sqrt{\mu}\int_{-\mu}^{0}\frac{\sqrt{1+\frac{s}{\mu}}}{1+e^{-\beta s}}dt=O(\mu^{1/2})\]
 It gives formula (\ref{1int}) since\[
\int_{0}^{\mu}\frac{r(t)}{1+e^{\beta(\mu-t)}}dt=O(1)\]
 Let us show now that \begin{equation}
\int_{\mu}^{\infty}\frac{F_{\Lambda}^{(q)}(t)}{1+e^{\beta(t-\mu)}}dt=O(\mu^{1/2})\label{2int}\end{equation}
 After change of variable we get\[
\int_{\mu}^{\infty}\frac{F_{\Lambda}^{(q)}(t)}{1+e^{\beta(t-\mu)}}dt=\int_{0}^{\infty}\frac{F_{\Lambda}^{(q)}(\mu+t)}{1+e^{\beta t}}dt=\]
\[
=c\int_{0}^{\infty}\frac{\sqrt{\mu+t}}{1+e^{\beta t}}dt+\int_{0}^{\infty}\frac{r(\mu+t)}{1+e^{\beta t}}dt\]
 It is clear that\begin{align*}
\int_{0}^{\infty}\frac{\sqrt{\mu+t}}{1+e^{\beta t}}dt & =\sqrt{\mu}\int_{0}^{\infty}\frac{\sqrt{1+\frac{t}{\mu}}}{1+e^{\beta t}}dt=O(\mu^{1/2})\\
\int_{0}^{\infty}\frac{r(\mu+t)}{1+e^{\beta t}}dt & =O(1)\end{align*}
 Formulas (\ref{0int}), (\ref{1int}) and (\ref{2int}) prove the
lemma.

Now the assertion of the theorem follows from (\ref{equiv}).

\paragraph{Proof of theorem 3}

By theorem 2 we have\begin{equation}
\frac{1}{\Lambda}\ln\Xi_{\mu-C}^{(0)}\left(\Lambda\right)\leq\frac{1}{\Lambda}\ln\Xi_{\mu}^{(V)}\left(\Lambda\right)\leq\frac{1}{\Lambda}\ln\Xi_{\mu+C}^{(0)}\left(\Lambda\right)\label{ner}\end{equation}
 It is well known that \[
\lim_{\Lambda\rightarrow\infty}\frac{1}{\Lambda}\ln\Xi_{\mu}^{(0)}\left(\Lambda\right)=\frac{1}{\pi}\int_{0}^{\infty}\ln(1+e^{\beta(\mu-p^{2})})dp\]
 After the change of variables $t=p^{2\text{ }}$ and integrating
by parts we get\[
\int_{0}^{\infty}\ln(1+e^{\beta(\mu-p^{2})})dp=\int_{0}^{\infty}\frac{\sqrt{t}}{1+e^{\beta(t-\mu)}}dt\]
 It was shown above \[
\int_{0}^{\infty}\frac{\sqrt{t}}{1+e^{\beta(t-\mu)}}dt\sim K\mu^{\frac{3}{2}},\mu\rightarrow\infty\]
 So the result follows from (\ref{ner}).

\end{document}